\documentclass[prl,superscriptaddress,amsmath,amssymb,showpacs,noshowkeys,a4paper,twocolumn]{revtex4-1}

\usepackage{graphicx}
\usepackage{dcolumn}
\usepackage{bm}
\usepackage{epstopdf}
\usepackage{wasysym}
\usepackage{ulem}

\newcommand{\affpmmh}{\affiliation{Laboratoire de Physique et M\'ecanique des Milieux H\'et\'erog\`enes (PMMH), UMR
CNRS 7636 ; PSL - ESPCI, 10 rue Vauquelin, 75005 Paris, France; Sorbonne
Universit\'e - UPMC, Univ. Paris 06; Sorbonne Paris Cit\'e - UDD, Univ. Paris 07}}

\newcommand{\afflang}{\affiliation{ Institut Langevin, ESPCI, CNRS, PSL Research University, 6 rue Jussieu, 75005, Paris, France}}

\begin{document}
\title{Dispersion free control of hydroelastic waves down to sub-wavelength scale}

\author{L.~Domino}
\affpmmh
\author{M.~Fermigier}
\affpmmh
\author{E.~Fort}
\afflang
\author{A.~Eddi}
\affpmmh

\begin{abstract}
Hydroelastic surface waves propagate at the surface of water covered by a thin elastic sheet and can be directly measured with accurate space and time resolution. We present an experimental approach using hydroelastic waves that allows us to control waves down to the sub-wavelength scale. We tune the wave dispersion relation by varying locally the properties of the elastic cover and we introduce a local index contrast. This index contrast is independent of the frequency leading to a dispersion-free Snell-Descartes law for hydroelastic waves. We then show experimental evidence of broadband focusing, reflection and refraction of the waves. We also investigate the limits of diffraction through the example of a macroscopic analog to optical nanojets, revealing that any sub-wavelength configuration gives access to new features for surface waves.
\end{abstract}

\maketitle

%
%
Gravity-capillary waves have been extensively used as model waves to tackle the issue of wave control at macroscopic scale. Contrary to optics and acoustics, their temporal and spatial typical scales allow for direct and accurate observation of wave propagation inside the medium. The design and fabrication of media with given properties is generally obtained by tuning the local bathymetry, which modifies the wave phase velocity \cite{1978Lighthill}. Immersed structures have been used to obtain Anderson localization \cite{1988Belzons} or to create macroscopic metamaterials for wave focusing \cite{2005Hu,  2013Hu, 2014Wang, 2015Bobinski} and cloaking \cite{2008Farhat, 2013Berraquero}. Wave control is achieved for waves with wavelength larger or comparable to the liquid depth (shallow water approximation). In this regime, damping becomes a major issue in laboratory experiments as viscous friction at the bottom dissipates most of the mechanical wave energy. These two limitations narrow the effective bandwidth of the devices created with gravity-capillary waves.  

Here, we propose a novel approach based on hydroelastic waves \textit{i.e.} waves that propagate at the surface of water covered with an elastic sheet. These waves were initially introduced to describe motion in ice sheets located in the marginal ice zone \cite{1887Greenhill, 1985Davys, 1987Schulkes, 1988Squire} and later to study floating structures \cite{2004Watanabe}, wakes in the lubrication approximation \cite{2016Arutkin} or cloaking \cite{2016Zareei}. In the limit of thin membranes their dispersion relation writes
\begin{equation}
  \omega ^2 = \left (  g k + \frac{T}{\rho} k^3  +  \frac{D}{\rho}  k^5 \right ) \tanh{kh_0} , \\
  \label{eq:DispRel}
\end{equation}
where $\omega = 2 \pi f$ is the pulsation, $k = 2 \pi / \lambda$ is the wavenumber, $g=9.81$ m.s$^{-2}$ is the acceleration of gravity, $T$ is the mechanical tension in the elastic sheet, $\rho$ is the fluid density, $D$ is the flexural modulus of the elastic sheet and $h_0$ the fluid depth. The flexural modulus $D = \frac{Ee^3}{12(1-\nu^2)}$ depends on the Young's modulus $E$ of the material, its Poisson modulus $\nu$ and its thickness $e$. Eq. \ref{eq:DispRel} exhibits three distinct regimes depending on the material properties and the wave pulsation $\omega$: gravity waves, tension waves and flexural waves. So far, very few experiments at the laboratory scale highlighted the flexural regime using whether thin elastic polymer sheets \cite{2013Deike, 2013Montiel} or granular rafts \cite{2012Planchette}. 

Here, we propose to achieve spatial control of the propagation of hydroelastic waves by modifying the dispersion relation (eq. \ref{eq:DispRel}) through local variations of the sheet's flexural modulus $D$. We first describe our experimental set-up and verify quantitatively the prediction from eq. \ref{eq:DispRel}. We introduce a local index contrast using the local phase velocity. This index contrast is independent of the frequency, which allows us to define a dispersion-free Snell-Descartes law for hydroelastic waves. To show the versatility of the system we then implement more complex structures to focus wave energy and probe wave effects due to the finite size of the system.

\section{Experimental setup}
\begin{figure}
\includegraphics[width=.99\linewidth]{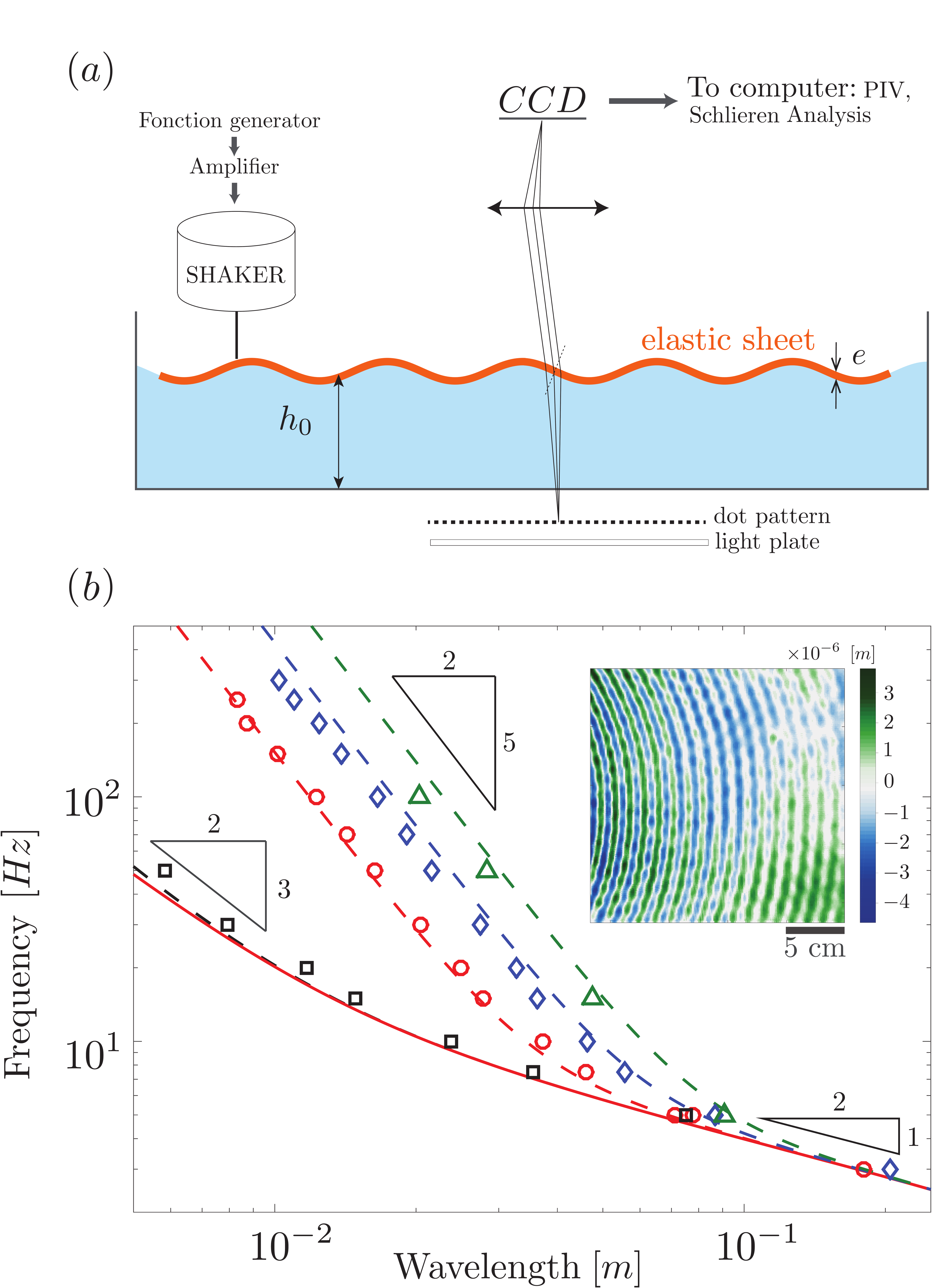}
  \caption{(a) Sketch of the experimental set-up showing the container filled with water and covered with an elastic sheet, the wave generation device and the imaging system. 
  (b) Measured dispersion relation for $4$ different film thicknesses 
  $e=20\ \mu$m (squares), 
  $300\ \mu$m (circles), 
  $500\ \mu$m (diamonds) and 
  $800 ~\mu$m (triangles). 
  For each thickness, the dashed line is the theoretical dispersion relations obtained using equation \ref{eq:DispRel}. 
  The plain red line shows the theoretical gravity-capillary dispersion relation for water waves. 
  Inset: Typical wave field measured for a point source vibrating at $100$ Hz. The wave travels from left to right. 
  }
\label{fig:Fig1}
\end{figure}

We use a glass tank ($80$ cm $\times$ $40$ cm $\times$ $20$ cm) filled to a depth $h_0=16.5$ cm of water. We cover its surface with a $75$ cm $\times$ $35$ cm wide elastic sheet of thickness $e=20$ --- $800$ $\mu$m made of an optically transparent silicone rubber sheet with Young's modulus $E=1.47\pm0.09$ MPa, density $\rho_s = 970$kg/m$^3$ and Poisson's ratio $\nu = 0.5$. This elastic film floats freely at the surface of water so that the mechanical tension $T$ inside reduces to the water surface tension $T = \sigma = 50$ mN/m. The waves are generated with a vibration exciter powered with an amplifier controlled with a waveform generator. We work with frequencies ranging from $2$ to $200$ Hz and with amplitudes $\zeta \ll \lambda$ to ensure the waves are in the linear regime. In addition, to guarantee that we are in the thin membrane limit we check that $\rho_s e \omega^2 \ll \{ D k^4, T k^2, \rho g\} $, so that equation 1 is valid.

To analyze quantitatively the wave field we use the Free-Surface Synthetic Schlieren optical technique \cite{2009Moisy} based on the apparent displacement of a random dot pattern due to the local slope of the interface. The pattern is located underneath the tank and we observe it from the top using a CCD camera located at $H\simeq2$ m from the fluid surface [Fig. \ref{fig:Fig1}(a)]. The sampling frequency of the camera is set to obtain stroboscopic images of the wave propagation with at least 12 images per period. The area we observe with our camera is about $20$ cm $\times$ $20$ cm wide ($2048\times2048$ pixels). A Digital Image Correlation (DIC) algorithm (PIVlab \cite{2014Thielicke, 2014Thielicke2}) is used to compute the displacement field between each recorded image and the reference image. After reconstruction, we obtain 2D elevation fields ($255\times255$ points) and we are able to measure amplitudes down to $\zeta = 1 \ \mu $m. An example of the obtained circular height field generated by a point source is shown in [Fig. \ref{fig:Fig1}b] (inset).


\section{Dispersion relation}

We first probe the validity of the theoretical dispersion relation predicted by eq. (\ref{eq:DispRel}). We measure the wave field for circular waves at various frequencies in the range $f=2 $ --- $ 200$ Hz obtained with a point source vibrating on elastic films with thicknesses $e=20 $ --- $800$ $\mu$m. For each field, we perform 2D spatial Fourier transforms to determine the wavenumber $k$ associated to each frequency. We plot in figure \ref{fig:Fig1}(b) the forcing frequency $f$ against the measured wavelength $\lambda$ on a log-log scale. 
Our measurements show that the wavelength decreases with the forcing frequency $f$, typically ranging from $\lambda = 20$ cm to $\lambda = 0.5$ cm. For low frequencies ($f<5$ Hz) we observe a slope of $-1/2$ revealing that the gravity term $\omega^2 \simeq gk$ in eq. (\ref{eq:DispRel}) is dominant. For larger frequencies, two regimes can be observed. For thin elastic sheets ($e=20$ $\mu$m) we observe a slope of $-3/2$ which corresponds to the tension term $ T/\rho~k^3 = \sigma / \rho ~k^3 $. This experimental dispersion relation is in perfect agreement with theory. The transition between the gravity and the tension regime occurs for $\lambda = 2\pi \sqrt{T/ \rho g } = 1.4 \ 10^{-2}$ m. This part of the dispersion relation corresponds to standard water waves, confirming that tension in the film is solely due to the liquid surface tension. For thicker films the behavior is markedly different. The measured slope is $-5/2$ showing that flexural term $D/\rho ~k^5$ is leading. This is confirmed by the theoretical dispersion relation that is in excellent agreement with our experimental data. This flexural regime is reached when $k \geq \sqrt[4]{\rho g / D}$ and $k \geq \sqrt{T /D}$. In the following we will only consider waves in the flexural regime, \textit{i.e.} hydroelastic waves. 

\section{Effective index and broadband refraction}

\begin{figure*}[t]
  \centering
  \includegraphics[width=1\linewidth]{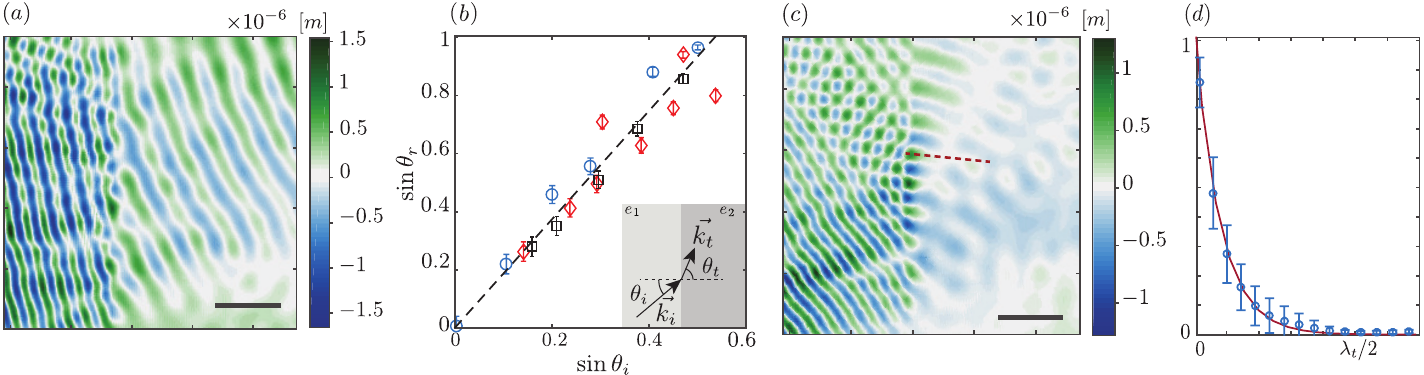}
  \caption{
(a) Wave field showing refraction at the interface between two media, with $e_1= 300$ $\mu$m (left) and $e_2=800$ $\mu$m (right). The wave travels from the left to the right. The scale bar represents  $5$ cm. 
(b) Sinus of the measured refracted angle $\theta_t$ as a function of the sinus of the incident angle $\theta_i$ for 3 different frequencies. The dashed line corresponds to Snell's law prediction with $n_1/n_2=1.81$. Inset: Schematic drawing of the experiment, showing the wave vectors and the angles of the incident wave $\theta_i$ and of the transmitted wave $\theta_t$. 
(c) Wave field showing the total reflection at the interface between the two media previously described. The wave travels from the left to the right. The scale bar represents  $5$ cm. 
(d) Blue circles: normalized profile of the intensity of the evanescent wave taken along the dashed line shown in (c). Red line: exponential fit.
}
\label{fig:Fig2}
\end{figure*}
 
In this regime the dispersion relation (eq. \ref{eq:DispRel}) in the deep water approximation can be simplified as 

\begin{equation}
  \omega ^2 \simeq    \frac{D}{\rho}  k^5. \\
  \label{eq:DispRel2}
\end{equation}

The phase velocity $v_\varphi = \frac{\omega}{k} = \sqrt{D/\rho }\ k^{3/2}$ then only depends on the film properties and the wavenumber $k$.
%
%
From this phase velocity we can define a relative effective refractive index $n(D, k) \propto \frac{1}{v_\varphi}$. $n$ is spatially tunable by varying locally the value of the flexural coefficient $D$. This can be achieved by changing locally the Young's modulus $E$ or the film thickness $e$.  For two domains covered with elastic films with different thicknesses $e_1$ and $e_2$ and same Young's modulus $E$, the ratio of their refractive indices $n_1$ and $n_2$ writes
\begin{equation}
  \frac{n_1}{n_2} =   \frac{v_{\varphi 2}}{v_{\varphi 1}} = \frac{k_2}{k_1}  = \left (\frac{D_2}{D_1} \right )^{1/5}  =  \left (\frac{e_2}{e_1} \right )^{3/5}  . \\
  \label{eq:n1n2}
\end{equation}
Thicker regions (resp. thinner) thus correspond to smaller (resp. higher) refractive indices. In the hydroelastic regime the index ratio is given by the local values of the film thickness $e$ and does not depend on $k$ nor $\omega$. 

We perform experiments to test eq. \ref{eq:n1n2} through Snell-Descartes law using the refraction of a plane wave at an interface between two media. The interface is obtained using two thicknesses ($e_1=300\ \mu$m and $e_2=800\ \mu$m, respectively) [Fig. \ref{fig:Fig2}(a)], and the frequency $f$ ranges from $50$---$200$ Hz. The incident (resp. transmitted) waves have a wave vector $\vec{k_i}$ (resp. $\vec{k_t}$) that forms an angle $\theta_i$ (resp. $\theta_t$) with the interface normal [Fig. \ref{fig:Fig2}b]. We measure these angles by means of spatial Fourier transforms for varying incidence angle, ranging from $0^{\circ}$ to $40^{\circ}$. We plot in figure \ref{fig:Fig2}(b) $\sin{\theta_t}$ against $\sin{\theta_i}$ for three different frequencies. The result is linear which means that hydroelastic waves obey the Snell-Descartes law of refraction: $n_1 \sin{\theta_i} = n_2 \sin{\theta_t}$. This result based on translation invariance holds whatever the frequency. The expected slope given by the refractive index ratio $ \frac{n_{1}}{n_{2}}   \simeq 1.81 $ is in excellent agreement with our experimental data.

We also study the situation where the angle of incidence is larger than the critical angle, here $\arcsin({1/1.8}) \simeq  33.7^{\circ}$. Such a wave field is presented in figure \ref{fig:Fig2}(c) where $\theta_i \simeq 40^{\circ}$. 
As expected, the wave undergoes a Total Internal Reflection (TIR) with the presence of an evanescent wave in medium 2. No energy is transmitted through the interface but this evanescent wave can be observed: the height of the wave decreases rapidly away from the interface. We can directly measure the amplitude of these waves on our fields, and we represent the profile obtained in the inset of figure  \ref{fig:Fig2}(d). The amplitude of these waves decreases exponentially, with a typical penetration length $\delta \simeq (2.1 \pm 0.5)\ 10^{-1} \times \lambda _t$. 
This penetration length is in good agreement with the expected value of $\delta_{th} = 1/\kappa = 2.6 \ 10^{-1} \times \lambda _t$ obtained with $\kappa = \omega/c \sqrt{(n_1 \sin{\theta_i})^2 - n_2^2}$.

\section{Lenses, focalisation}
\label{sec:Lens}

\begin{figure}[h]
  \centering
  \includegraphics[width=1\linewidth]{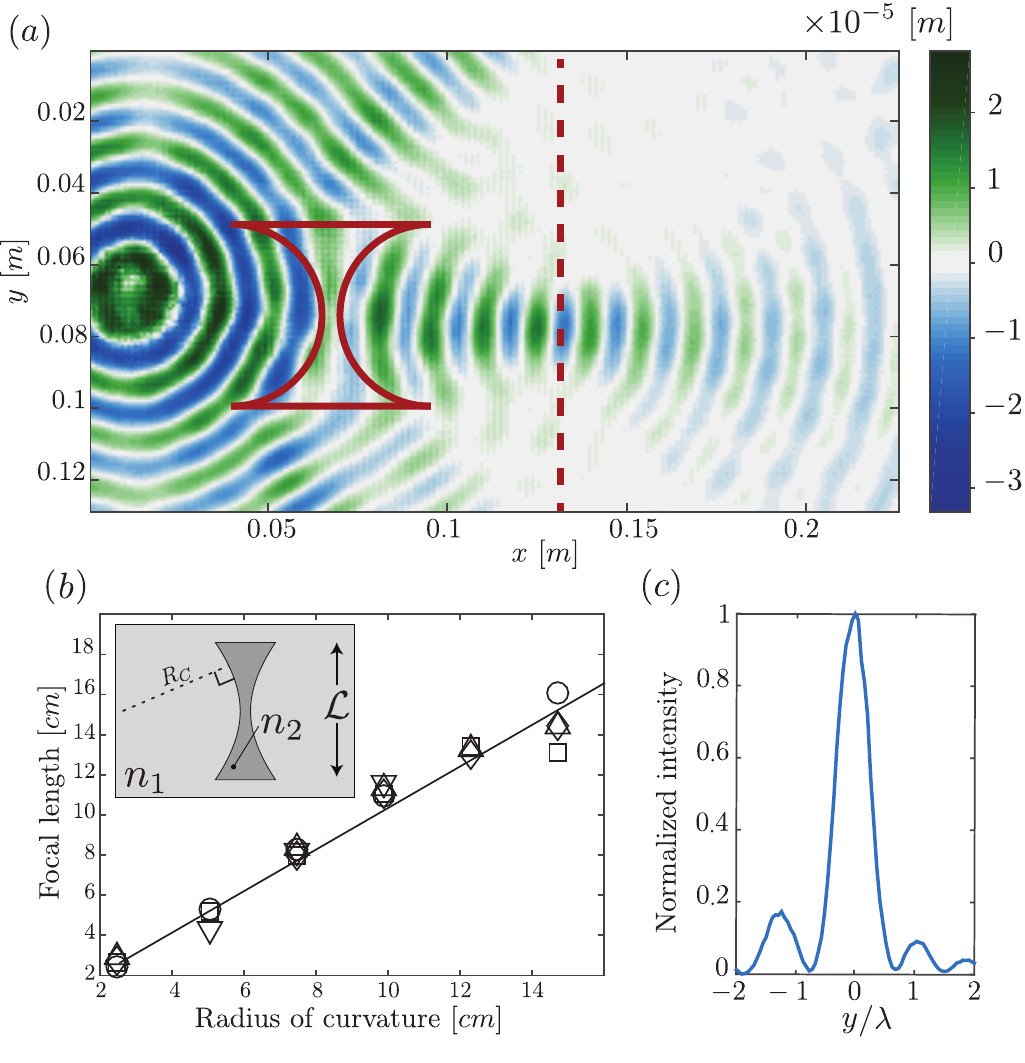}
  \caption{
    (a) Wave field for a lens with a radius of curvature of $2.5$ cm. Circular waves traveling at $f=75$ Hz from the left to the right. 
    The plain red line denotes the shape of the lens, the dashed line denotes its focal plane.
     (b) Measurement of $L_f$ as a function of $R_C$ for various lenses and 5 different frequencies: 
     $50$ Hz (squares), 
     $75$ Hz (diamonds), 
     $100$ Hz (up triangles), 
     $125$ Hz (down triangles) and 
     $150$ Hz (circles).
     The error on the measure is lower than $1\%$ and is much smaller than the marker size. Plain line denotes the theoretical focal length for a thin lens (see text).
     (c) Profile of the intensity field along the focal plane for a lens with radius $13.7$ cm and for $f = 50$ Hz.
     }
\label{fig:Fig3}
\end{figure}
Fine sub-wavelength wave control can be achieved easily by designing engineered shapes to focus and guide waves. Here, as an example, we design 2D lenses by cutting out symmetric circular arcs in the silicon polymer. The obtained shapes [Figure \ref{fig:Fig3}] are defined by their radius of curvature $R_C$. 
We then deposit them on the first membrane to locally increase the thickness. Note that this shape should create a convergent lens as thicker regions have a lower refractive index $n_2<n_1$. We excite the system using a point source located on the left of the lens. Fig. \ref{fig:Fig3}(a) presents a typical wave field for a lens with $R_C=2.5$ cm excited with $f=75$ Hz and shows a focal spot on the right side of the lens (see also supplementary videos: Movie1.mov and Movie2.mov). Using the location of this focal spot and that of the source we define the focal length $L_f$ as $1/L_f = 1/s + 1/s'$, where $s$ (resp. $s'$) is the distance between the lens and the source (resp. the image). [Fig. \ref{fig:Fig3}(b)] shows the measured $ L_f$ as a function of $R_C$ for $5$ different frequencies ranging from $50$ Hz to $150$ Hz. We observe that $L_f$ increases linearly with $R_C$ while being independent of $f$ as the refractive index ratio only depends on the film thickness ratio in the flexural regime.
Here the typical size $\mathcal{L}$ of the lens compares with $\lambda$ and the paraxial approximation is clearly not satisfied. This makes ray optics a poor candidate to model our results. However, its prediction $ L_f = \frac{R_C}{2 ( 1 - \frac{n_2}{n_1} )}$ agrees surprisingly well with our experimental data [Fig. \ref{fig:Fig3}(b)].

We now characterize the profile of the wave field at the focus. Fig. \ref{fig:Fig3} (c) presents the lateral normalized intensity profile of the focal spot obtained for a lens with a curvature radius of $R_C = 13.7$ cm at $f = 50$ Hz. The profile exhibits a central peak with a Full Width at Half Maximum (FWHM) of $0.63\lambda$, and 2 side lobes. 
The presence of these secondary peaks is the signature of diffraction: as both the typical width of the focal spot and the typical size of the lens $\mathcal{L}$ compare with $\lambda$, wave propagation should be described at the wavelength scale.



\section{Sub-wavelength focusing}
\label{sec:FourierLens}

\begin{figure}[h]
  \centering
  \includegraphics[width=1\linewidth]{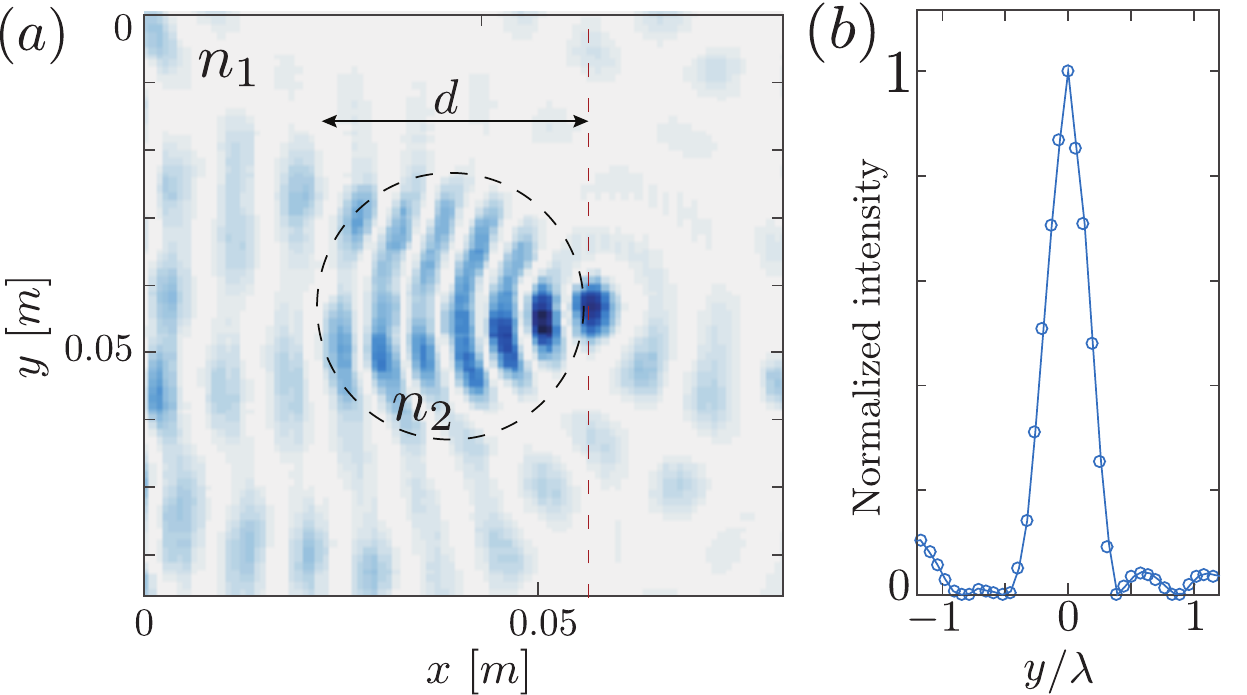}
  \caption{ 
(a) Intensity profile for a plane wave propagating at $150$ Hz through a thin disk of diameter $d$, here denoted with the dark dashed line. The film is $300$ $\mu$m thick inside the circle, and $800$ $\mu$m thick elsewhere. We take an intensity profile along the red vertical dashed line. 
(b) Intensity profile taken along the red dashed line plotted in (a).
}
\label{fig:Fig4}
\end{figure}
To further show the versatility of hydroelastic wave to control waves at the subwavelength scale, we use a geometry known in optics as ``nanojet''. They were first introduced in optics using small cylindrical (or spherical) structures ($\mathcal{L}\sim 10 \lambda$) with a strong index contrast \cite{2004Chen, 2008Ferrand, 2010Geints}. These structures are traditionally manufactured or simulated using a dielectric sphere that has a higher refractive index than the surrounding medium, like glass, water or latex. The focal spot is created by the combination of evanescent and propagating waves in the shadow-side of the sphere.

We transpose this object to 2D hydroelastic waves by creating a thinner circular area with diameter $d$ in the elastic sheet [Fig. \ref{fig:Fig4}(a)]. As for classical microspheres in optics, the refractive index $n_2$ in this region is higher than $n_1$ in the outside medium. [Fig. \ref{fig:Fig4}(a)] shows the measured intensity field inside and outside the disk for an incident wave with $f=150~$Hz and $d=32~$mm (see also supplementary videos Movie1.mov and Movie2.mov). The wave propagates from the left to the right, and the thinner region is denoted with the dark dashed circle. We observe that the circular patch distorts the incoming plane wave and that a strong focal spot emerges out of the circular area. We show in Fig. \ref{fig:Fig4}(b) the lateral intensity profile of this focal spot that presents a very narrow peak as well as small side lobes on both sides. The FWHM of this focal spot is $0.33 \lambda$, which is smaller than the classical diffraction limit at $0.5\lambda$, confirming that sub-wavelength focusing can be achieved using this simple design at the wavelength scale.

\section{Conclusion}
 
We have achieved control of hydroelastic waves propagation in a model experiment. We first confirmed that the waves can be accurately described by the dispersion relation (Eq. \ref{eq:DispRel}) at the laboratory scale, and that the elastic sheet's properties have a crucial incidence on the wave propagation in the flexural waves regime. Indeed, the flexion modulus $D$ can be tuned spatially by locally modifying the elastic film's thickness. We build a local index contrast that only depends on the film's flexion modulus $D$ and is therefore dispersion free despite the dispersive nature of the waves. Using this we first show that Fermat principle applies for hydroelastic waves, as refraction of an incident plane wave on a straigth interface obeys Snell-Descartes law independently of the incident wave frequency. With this feature we implement lenses with tunable focalisation properties. Nevertheless, the index variations occur on a typical scale that compares with the hydroelastic wavelength leading to subtle wave effects. This is particularly revealed by the construction of a macroscopic equivalent of nanojets. These simple circular structures allow to overcome the diffraction limit leading to a focal spot as small as $\lambda/3$.


We believe our macroscopic experiment can be used as a model experiment to study the physics of waves with new features so far unachieved. Direct observation of the waves combined with the ability to tune the medium's properties down to sub-wavelength scale opens promising perspectives to probe wave propagation in structured \cite{1996Krauss, 2001Vasseur, 2006Pendry, 2014Brule} or random media  \cite{1993Duvall, 2010Vellekoop, 2012Mosk}. In particular this system allows for precise spatio-temporal control of wave sources as well as a precise monitoring of dynamical effects. Inspired by recent work on time-reversal of gravito-capillary waves \cite{2016Bacot}, we now aim to implement macroscopic dynamical spatial structures \cite{2010Chumak, 2011Sivan, 2016Yuan}.

\end{document}